\documentclass{sig-alternate}

\DeclareMathAlphabet{\mathitbf}{OML}{cmm}{b}{it}

\begin{document}

\title{Building a Balanced k-d Tree with MapReduce} 

%\subtitle{}

\author{Russell A. Brown}

\date{16 December 2015}
\maketitle
\begin{abstract}

The original description of the \emph{k}-d tree recognized that rebalancing techniques, such as are used to build an AVL tree or a red-black tree, are not applicable to a \emph{k}-d tree.  Hence, in order to build a balanced \emph{k}-d tree, it is necessary to obtain all of the data prior to building the tree then to build the tree via recursive subdivision of the data.  One algorithm for building a balanced \emph{k}-d tree finds the median of the data for each recursive subdivision of the data and builds the tree in $O\left(n \log n\right)$ time.  A new algorithm builds a balanced \emph{k}-d tree by presorting the data in each of $k$ dimensions prior to building the tree, then preserves the order of the $k$ presorts during recursive subdivision of the data and builds the tree in $O\left(kn \log n\right)$ time.  This new algorithm is amenable to execution via MapReduce and permits building and searching a \emph{k}-d tree that is represented as a distributed graph.

\end{abstract}

%\category{C.4}{Performance of systems}{Measurement techniques, Modeling techniques}
%\category{D.1.3}{Software Engineering}{Concurrent Programming}[Parallel programming]

%\terms{terms here ...}

%\keywords{keywords here ...}

\section{Introduction}
\label{sec:introduction}
Bentley introduced the \emph{k}-d tree as a binary tree that stores \emph{k}-dimensional data \cite{Bentley}.  Like a standard binary tree, the \emph{k}-d tree subdivides data at each recursive level of the tree.  Unlike a standard binary tree that uses only one key for all levels of the tree, the \emph{k}-d tree uses $k$ keys and cycles through these keys for successive levels of the tree.  For example, to build a \emph{k}-d tree from two-dimensional points that comprise $\left(x,y\right)$ coordinates, the keys would be cycled as $x,y,x,y,x,y...$ for alternate levels of the \emph{k}-d tree.

Due to the use of different keys at successive levels of the tree, it is not possible to employ rebalancing techniques, such as are used to build an AVL tree \cite{Adelson} or a red-black tree \cite{Bayer,Guibas}, when building a \emph{k}-d tree.  Hence, the typical approach to building a balanced \emph{k}-d tree finds the median of the data for each recursive subdivision of those data.  Bentley showed that if the median of $n$ elements were found in $O\left(n\right)$ time, it would be possible to build a depth-balanced \emph{k}-d tree in $O\left(n \log n\right)$ time.  However, algorithms that find the median in guaranteed $O\left(n\right)$ time are somewhat complicated \cite{Blum,Cormen}.  An alternative approach to building a balanced \emph{k}-d tree presorts the data in each of $k$ dimensions prior to building the tree using merge sort \cite{Neumann,Knuth,Knuth2} or heap sort \cite{Williams}.  The order of the $k$ presorts is then maintained when building a balanced tree and thereby permits a worst-case complexity of $O\left(kn \log n\right)$ for building the tree \cite{Brown}.

The remainder of this article will use as an example a \emph{k}-d tree that sorts rectangular bounding boxes \cite{Sproull}.  This \emph{k}-d tree permits a directed search of the tree by a query bounding box in order to determine which of the bounding boxes in the tree the query bounding box intersects.  The principles of building and searching such a \emph{k}-d tree will be discussed initially for a tree that resides in memory, then extended to a \emph{k}-d tree that is represented as a graph whose elements are distributed across multiple MapReduce compute nodes \cite{Lin}.

\section{Building a Memory-Resident Tree}
\label{sec:build_memory_kd_tree}

\begin{figure}[h]
\centering
\centerline{\includegraphics[width=3.5in]{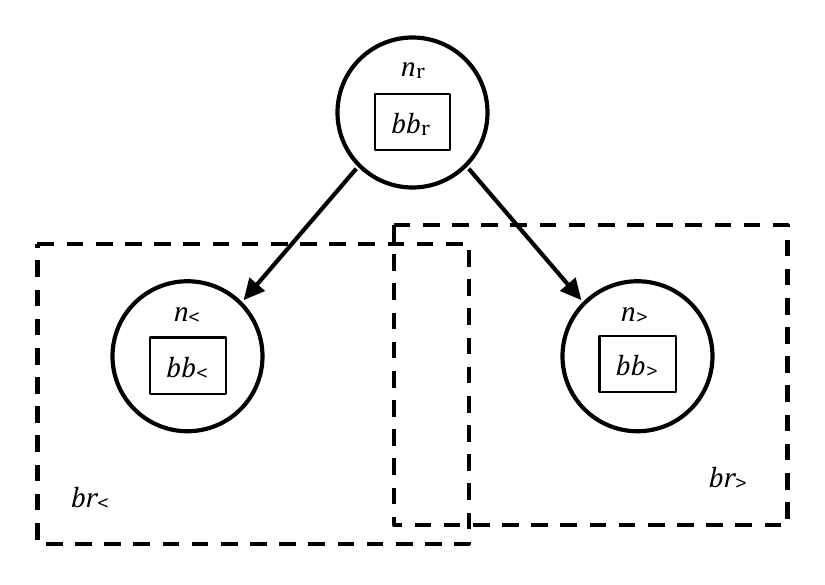}}
\caption{The root node $n_\mathrm{r}$ of the \emph{k}-d tree and its two children $n_\mathrm{<}$ and $n_\mathrm{>}$ are depicted by circles.  The root node stores the bounding box $bb_\mathrm{r}$ and the bounding regions $br_\mathrm{<}$ and $br_\mathrm{>}$ that are depicted by dashed rectangles whose sizes and locations are not represented accurately in this figure.}
\label{fig:RootNodeOfTree}
\end{figure}

The \emph{k}-d tree will be used to search for bounding boxes that intersect a query bounding box.  Hence, each node of the tree must store the $\left(x_\mathrm{min}, y_\mathrm{min}, x_\mathrm{max}, y_\mathrm{max}\right)$ coordinates of a bounding box to facilitate an intersection test of that bounding box against the query bounding box, as shown for the root node in Figure \ref{fig:RootNodeOfTree}.  In addition, each node of the tree must permit a determination of which of the node's two subtrees the query bounding box will search recursively.  This determination is facilitated by storing, for each subtree, the $\left(x_\mathrm{min}, y_\mathrm{min}, x_\mathrm{max}, y_\mathrm{max}\right)$ coordinates of a rectangular bounding region that just encloses all of the bounding boxes in the subtree.  If the query bounding box intersects a bounding region, the query bounding box must search the associated subtree.

Given the above requirements for each node of the \emph{k}-d tree, the tree may be constructed recursively as summarized below.  A detailed description of the k-d tree building algorithm has been published previously \cite{Brown}.

First, the bounding boxes are presorted independently in $x_\mathrm{min}$ and $y_\mathrm{min}$ prior to building the \emph{k}-d tree to create two presorted arrays of bounding boxes.  In order that each bounding box may have a unique sorting key in each of $x_\mathrm{min}$ and $y_\mathrm{min}$, a unique name $n_\mathrm{i}$ is assigned to each bounding box and used to form the super keys $x_\mathrm{min}$:$n_\mathrm{i}$ and $y_\mathrm{min}$:$n_\mathrm{i}$ for use in the presorts, where the colon represents the catenation operator.

\begin{figure}[h]
\centering
\centerline{\includegraphics[width=3.7in]{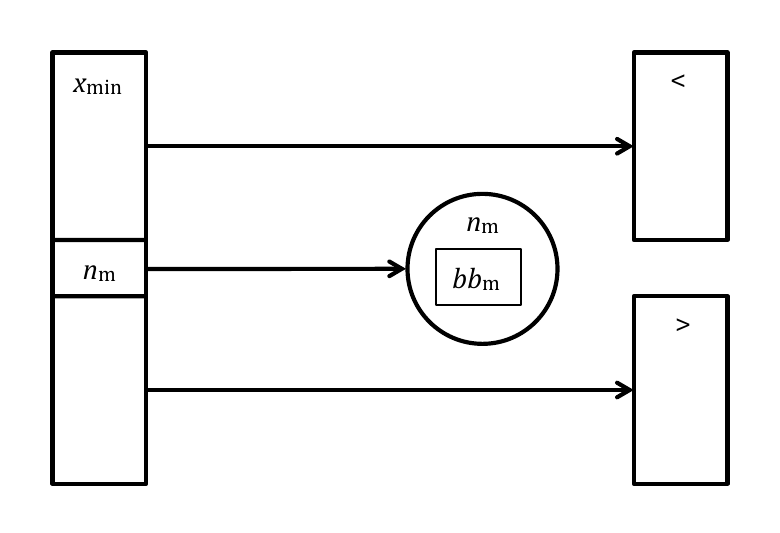}}
\caption{The $x_\mathrm{min}$:$n_\mathrm{i}$-sorted array ($x_\mathrm{min}$) is partitioned trivially at its median element ($n_\mathrm{m}$) to obtain the ``less than" $x_\mathrm{min}$:$n_\mathrm{i}$-sorted subarray ($<$) and the ``greater than" $x_\mathrm{min}$:$n_\mathrm{i}$-sorted subarray ($>$).  The bounding box $bb_\mathrm{m}$ of the median element is stored at the root of the tree that is depicted by the circle.}
\label{fig:TrivialPartition}
\end{figure}

Next, the presorted arrays are partitioned in $x_\mathrm{min}$ at the root of the tree as follows.  The bounding box $bb_\mathrm{m}$ at the median element $n_\mathrm{m}$ of the array that was presorted by the $x_\mathrm{min}$:$n_\mathrm{i}$ super key is stored at the root of the tree.  This median element trivially partitions the $x_\mathrm{min}$:$n_\mathrm{i}$-sorted array as shown in Figure \ref{fig:TrivialPartition}.  The elements of the $x_\mathrm{min}$:$n_\mathrm{i}$-sorted array at all addresses below the median address form a ``less than" $x_\mathrm{min}$:$n_\mathrm{i}$-sorted subarray.  The elements of the $x_\mathrm{min}$:$n_\mathrm{i}$-sorted array at all addresses above the median address form a ``greater than" $x_\mathrm{min}$:$n_\mathrm{i}$-sorted subarray.

\begin{figure}[h]
\centering
\centerline{\includegraphics[width=3.6in]{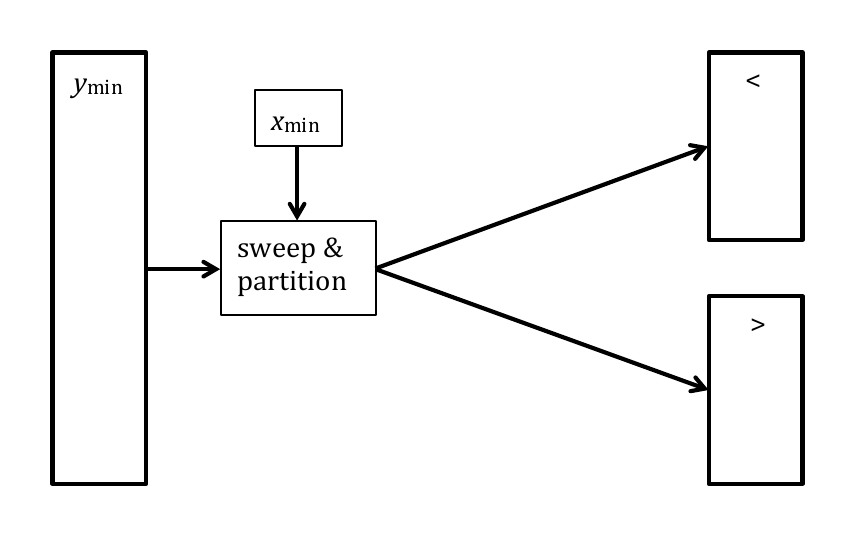}}
\caption{The $y_\mathrm{min}$:$n_\mathrm{i}$-sorted array ($y_\mathrm{min}$) is swept from lowest address to highest address and partitioned by the $x_\mathrm{min}$ of the median element $n_\mathrm{m}$ (see Figure \ref{fig:TrivialPartition}) of the $x_\mathrm{min}$:$n_\mathrm{i}$-sorted array to obtain the ``less than" $y_\mathrm{min}$:$n_\mathrm{i}$-sorted subarray ($<$) and the ``greater than" $y_\mathrm{min}$:$n_\mathrm{i}$-sorted subarray ($>$).}
\label{fig:SweepAndPartition}
\end{figure}

This median element also partitions the $y_\mathrm{min}$:$n_\mathrm{i}$-sorted array, albeit non-trivially, via a ``sweep and partition" algorithm as shown in Figure \ref{fig:SweepAndPartition} \cite{Wald,Brown}.  In order to partition the $y_\mathrm{min}$:$n_\mathrm{i}$-sorted array, the array is swept from lowest to highest address, \emph{i.e.}, swept in order from lowest to highest $y_\mathrm{min}$:$n_\mathrm{i}$ super key, and the elements of this array are partitioned into a ``less than" $y_\mathrm{min}$:$n_\mathrm{i}$-sorted subarray and a ``greater than" $y_\mathrm{min}$:$n_\mathrm{i}$-sorted subarray as follows.  Each element's $x_\mathrm{min}$:$n_\mathrm{i}$ super key is compared to the $x_\mathrm{min}$:$n_\mathrm{m}$ super key of the median of the $x_\mathrm{min}$:$n_\mathrm{i}$-sorted array whose bounding box is stored at the root of the tree.  If the element's super key is less than the root's super key, the element is partitioned into the ``less than" $y_\mathrm{min}$:$n_\mathrm{i}$-sorted array.  If the element's super key is greater than the root's super key, the element is partitioned into the ``greater than" $y_\mathrm{min}$:$n_\mathrm{i}$-sorted array.  If the element's super key equals the root's super key, the element is ignored.  In this manner, the $y_\mathrm{min}$:$n_\mathrm{i}$-sorted array is partitioned by the $x_\mathrm{min}$:$n_\mathrm{m}$ super key of the root of the tree to create ``less than" and ``greater than" $y_\mathrm{min}$:$n_\mathrm{i}$-sorted subarrays that each preserve the relative $y_\mathrm{min}$:$n_\mathrm{i}$-sorted order from the initial presort.

The ``less than" $x_\mathrm{min}$:$n_\mathrm{i}$-sorted and $y_\mathrm{min}$:$n_\mathrm{i}$-sorted subarrays are used to build recursively the ``less than" subtree at the next level of the nascent \emph{k}-d tree.  The ``greater than" $x_\mathrm{min}$:$n_\mathrm{i}$-sorted and $y_\mathrm{min}$:$n_\mathrm{i}$-sorted subarrays are used to build recursively the ``greater than" subtree at the next level of the nascent \emph{k}-d tree.  At the root of each of these two subtrees, the $y_\mathrm{min}$:$n_\mathrm{i}$-sorted subarray is partitioned trivially.  The $x_\mathrm{min}$:$n_\mathrm{i}$-sorted subarray is partitioned non-trivially via the ``sweep and partition" algorithm that compares $y_\mathrm{min}$:$n_\mathrm{i}$ super keys.  The alternation of trivial and non-trivial partitioning in $x_\mathrm{min}$ and $y_\mathrm{min}$ continues at each level of the \emph{k}-d tree until the subarrays are exhausted at the leaf nodes of the tree.

\begin{figure}[h]
\centering
\centerline{\includegraphics[width=3.5in]{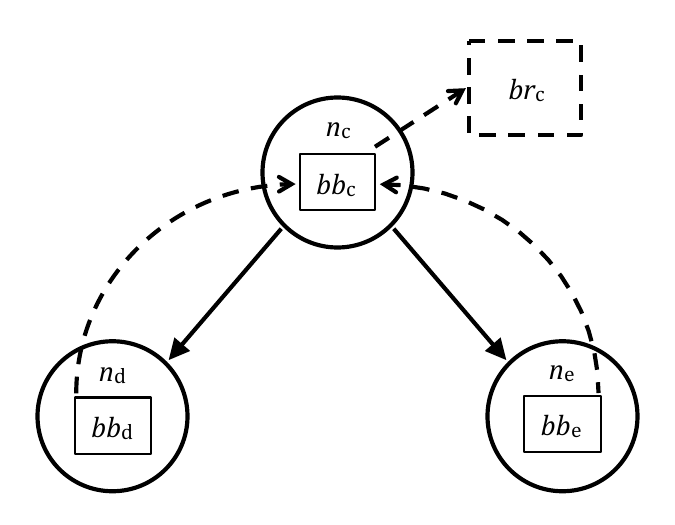}}
\caption{The bounding boxes $bb_\mathrm{d}$ and $bb_\mathrm{e}$ of the respective leaf nodes $n_\mathrm{d}$ and $n_\mathrm{e}$ are identical to the bounding regions of these nodes.  The bounding regions of the leaf nodes are combined with the bounding box $bb_\mathrm{c}$ of their parent node $n_\mathrm{c}$ as indicated by the curved, dashed arrows to calculate the bounding region $br_\mathrm{c}$ of the parent node that is depicted by the dashed rectangle.  Neither the size nor the location of $br_\mathrm{c}$ is represented accurately in this figure.}
\label{fig:LeafNodesOfTree}
\end{figure}

As the recursion unwinds, pointers to the node's children are returned to each node of the nascent \emph{k}-d tree.  In addition, the bounding region for each node is calculated as depicted in Figure \ref{fig:LeafNodesOfTree}.  For a leaf node, the dimensions of the bounding region are identical to those of the bounding box.  For a non-leaf node that has one or two children, the bounding region is the region that just encloses the bounding box of the node and the bounding regions of the children.  This bounding region is calculated as the minimum of the $x_\mathrm{min}$, the minimum of the $y_\mathrm{min}$, the maximum of the $x_\mathrm{max}$ and the maximum of the $y_\mathrm{max}$ of the node and of the node's children.

\section{Building a Distributed Tree}
\label{sec:build_distributed_kd_tree}

\begin{figure}[h]
\centering
\centerline{\includegraphics[width=3.8in]{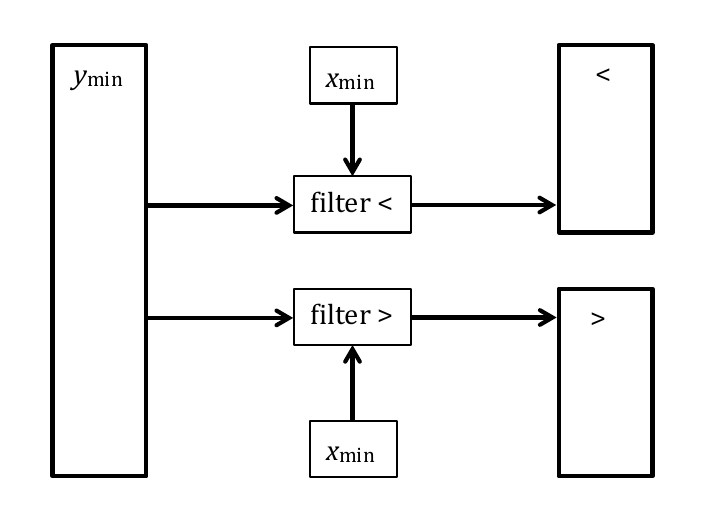}}
\caption{The $y_\mathrm{min}$:$n_\mathrm{i}$-sorted RDD ($y_\mathrm{min}$) is filtered by the $x_\mathrm{min}$ of the median element $n_\mathrm{m}$ (see Figure \ref{fig:SplitAt}) of the $x_\mathrm{min}$:$n_\mathrm{i}$-sorted RDD to obtain the ``less than" $y_\mathrm{min}$:$n_\mathrm{i}$-sorted RDD ($<$) and the ``greater than" $y_\mathrm{min}$:$n_\mathrm{i}$-sorted RDD ($>$).}
\label{fig:Filter}
\end{figure}

A distributed \emph{k}-d tree may be constructed with MapReduce using Spark.  The construction proceeds recursively in a similar manner to the construction of a memory-resident \emph{k}-d tree; however, arrays are not used to store the presorted bounding boxes.  Instead, Spark uses a \emph{resilient distributed dataset} (RDD), which represents a collection of elements that are distributed across multiple compute nodes of a cluster and hence may be processed in parallel \cite{KarauC}.  Spark provides many methods for processing RDDs in parallel.  Of particular interest are the sortByKey method \cite{KarauA} that may be used to presort an RDD prior to building the \emph{k}-d tree and the filter method \cite{KarauB} that may be used to ``sweep and partition" an RDD as shown in Figure \ref{fig:Filter}.

\begin{figure}[h]
\centering
\centerline{\includegraphics[width=3.6in]{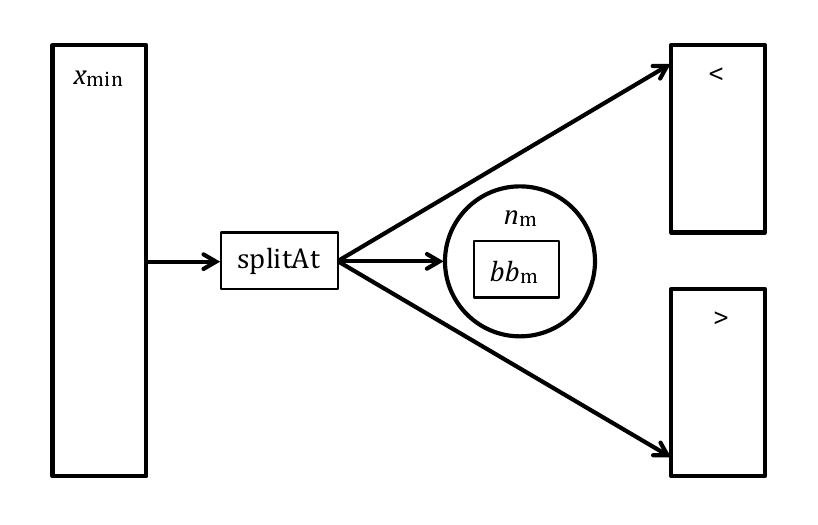}}
\caption{The splitAt method finds the median element ($n_\mathrm{m}$) of the $x_\mathrm{min}$:$n_\mathrm{i}$-sorted RDD ($x_\mathrm{min}$) then partitions the RDD at the median element to obtain the ``less than" $x_\mathrm{min}$:$n_\mathrm{i}$-sorted RDD ($<$) and the ``greater than" $x_\mathrm{min}$:$n_\mathrm{i}$-sorted RDD ($>$).  The bounding box $bb_\mathrm{m}$ of the median element is stored at the root of the tree that is depicted by the circle.}
\label{fig:SplitAt}
\end{figure}

Unfortunately, Spark provides neither a method that obtains the median element of a sorted RDD nor a method that splits a sorted RDD about its median element to create ``less than" and ``greater than" RDDs.  However, a proposed extension to Spark that implements the drop method, which deletes a specified number of elements from an RDD, provides insight into how the necessary splitting method may be implemented \cite{Erlandson}.  This splitting method, named splitAt, has been implemented to iterate over the partitions of an RDD in parallel in order to count the elements in each partition \cite{Zecevic} and thereby determine in which partition the median element lies and then obtain the median element of the RDD.  This median element is  used to split the RDD to create ``less than" and ``greater than" RDDs, as shown in Figure \ref{fig:SplitAt}.

The Spark sortByKey, filter and splitAt methods permit the construction of a distributed \emph{k}-d tree with MapReduce.  In contrast to a memory-resident tree, a distributed tree does not use pointers to link each \emph{k}-d node to its children.  Instead, the tree is represented as a graph whose nodes are distributed across multiple compute nodes as an RDD \cite{Lin}.  For example, the root node of the \emph{k}-d tree is represented as the following element of an RDD that contains all the nodes of the \emph{k}-d tree$$\left(n_{r}, \left(bb_\mathrm{r}, n_\mathrm{<}, br_\mathrm{<}, n_\mathrm{>}, br_\mathrm{>} \right) \right)$$where $n_\mathrm{r}$ represents the unique name of the root node of the tree, $bb_{r}$ represents the bounding box that is stored at the root node of the tree, $n_\mathrm{<}$ represents the unique name of the ``less than" child of the root node, $br_\mathrm{<}$ represents the bounding region that encloses the ``less than" subtree, $n_\mathrm{>}$ represents the unique name of the ``greater than" child of the root node, and $br_\mathrm{>}$ represents the bounding region that encloses the ``greater than" subtree.  This RDD will be named the \textbf{tree RDD} (in the following discussion, the name of a specific RDD will be designated in bold face text).

The parentheses in the above representation of the root node of the \emph{k}-d tree enclose \emph{tuples} \cite{Odersky} and in particular, the outer set of parentheses encloses a two-element tuple or pair whose first element is $n_\mathrm{r}$ and whose second element is the tuple $\left(bb_\mathrm{r}, n_\mathrm{<}, br_\mathrm{<}, n_\mathrm{>}, br_\mathrm{>} \right)$.  An RDD that comprises pairs is a \emph{pair RDD} and has the special property that the two elements of each pair function as a $\left( \mathrm{key, value} \right)$ pair.  In the case of the above-described root node of the \emph{k}-d tree, the unique name $n_\mathrm{r}$ is the key and the tuple $\left(bb_\mathrm{r}, n_\mathrm{<}, br_\mathrm{<}, n_\mathrm{>}, br_\mathrm{>} \right)$ is the value.  The $\left( \mathrm{key, value} \right)$ properties of a pair RDD permit Spark to retrieve a particular node of the \emph{k}-d tree from the \textbf{tree RDD} using the unique name of that node, analogous to the manner in which a pointer to a particular node of a memory-resident \emph{k}-d tree permits the retrieval of that node from the tree.

\vspace{0.5in}

\section{Searching a Distributed Tree}
\label{sec:search_distributed_kd_tree}

Searching a distributed \emph{k}-d is not performed in a recursive, depth-first manner such as is used to search a memory-resident tree.  Instead, an iterative, bread-first search is conducted in parallel by multiple bounding boxes via MapReduce, similar to the manner in which MapReduce executes Dijkstra's algorithm \cite{Lin}.  An overview of the search algorithm is presented below.

A search of the \emph{k}-d tree assumes the existence of a \textbf{search RDD} that contains the elements $\left(n_\mathrm{i}, bb_\mathrm{i} \right)$, where $bb_\mathrm{i}$ represents the bounding boxes that will search the tree for intersections and $n_\mathrm{i}$ represents the unique names of the bounding boxes.  In preparation for the first iteration of the search algorithm, the \textbf{search RDD} is processed using the Spark map method \cite{KarauB} to create a \textbf{query RDD} whose elements are $\left(n_\mathrm{r}, \left(n_\mathrm{i}, bb_\mathrm{i} \right) \right)$.  For each element of this \textbf{query RDD}, the key is $n_\mathrm{r}$ and represents the unique name of the root node of the \emph{k}-d tree.  This key specifies that each element of the \textbf{search RDD} will visit the root of the \emph{k}-d tree.  The \textbf{query RDD} is joined to the \textbf{tree RDD} using the Spark join method \cite{KarauA} that performs an \emph{inner join} and returns a \textbf{visit RDD} that contains the elements$$\left(n_\mathrm{r}, \left( \left(n_\mathrm{i}, bb_\mathrm{i} \right), \left(bb_\mathrm{r}, n_\mathrm{<}, br_\mathrm{<}, n_\mathrm{>}, br_\mathrm{>} \right) \right) \right)$$The \textbf{visit RDD} represents the fact that each element of the \textbf{search RDD} visits the root of the \emph{k}-d tree.

The \textbf{visit RDD} is subsequently processed using the Spark flatMapValues method \cite{KarauA} in order to check for intersection between each $bb_\mathrm{i}$ and $bb_\mathrm{r}$.  Each intersection creates an element $\left(n_\mathrm{j}, n_\mathrm{r} \right)$ in an \textbf{intersection RDD} that represents the fact that the bounding boxes $bb_\mathrm{j}$ intersect the bounding box $bb_\mathrm{r}$, where $bb_\mathrm{j}$ is a subset of $bb_\mathrm{i}$.

The \textbf{visit RDD} is processed once again using the flatMapValues method in order to check for intersection between each $bb_\mathrm{i}$ and $br_\mathrm{<}$ as well as to check for intersection between each $bb_\mathrm{i}$ and $br_\mathrm{>}$.  Each intersection between $bb_\mathrm{i}$ and $br_\mathrm{<}$ creates an element $\left(n_\mathrm{<}, \left(n_\mathrm{k}, bb_\mathrm{k} \right) \right)$ in the \textbf{next query RDD}.  Similarly, each intersection between $bb_\mathrm{i}$ and $br_\mathrm{>}$ creates an element $\left(n_\mathrm{>}, \left(n_\mathrm{l}, bb_\mathrm{l} \right) \right)$ in the \textbf{next query RDD}.  The bounding boxes $bb_\mathrm{k}$ and $bb_\mathrm{l}$ are the subsets of $bb_\mathrm{i}$ that will visit the ``less than" and ``greater than" subtrees, respectively, during the next iteration of the search algorithm.  The iterative search algorithm terminates when the \textbf{next visit RDD} is empty, as will be the case when no intersections between $bb_\mathrm{i}$ and $br_\mathrm{<}$ or $br_\mathrm{>}$ are detected.

The \textbf{intersection RDD} from each iteration of the search algorithm is accumulated to a \textbf{cumulative intersection RDD} using the Spark union method \cite{KarauA}.  Following the final iteration of the algorithm, the $\left(n_\mathrm{s}, n_\mathrm{t} \right)$ pairs of the \textbf{cumulative intersection RDD} are reorganized using the Spark groupByKey method \cite{KarauA} to create the pairs $\left(n_\mathrm{s}, [n_\mathrm{u}...n_\mathrm{w}] \right)$, where for each pair, $[n_\mathrm{u}...n_\mathrm{w}]$ specifies the list of bounding boxes $[bb_\mathrm{u}...bb_\mathrm{w}]$ that are a subset of $bb_\mathrm{t}$ and that intersect $bb_\mathrm{s}$.

A deeper understanding of the distributed \emph{k}-d tree search algorithm may be obtained by studying the Scala source code and embedded comments that are included with this article.

\section{Performance}
\label{sec:performance}

Under the Spark execution model, a \emph{master} distributes data and computation across multiple compute nodes known as \emph{workers}. The master and workers may comprise individual compute nodes of a cluster or they may comprise separate cores of a CPU. For the experimental results reported below, the master and workers comprise separate cores of a 2.3 GHz Intel quad-core i7 CPU that supports concurrent execution of two threads per core.

A test data set comprises 16 rectangles contained within a square; nine of the 16 rectangles intersect one another. The square is replicated and translated in the $\left(x, y\right)$ plane to produce a specified number of adjacent but non-overlapping squares, each of which contains 16 rectangles. The complete set of rectangles from all of the squares is used to construct a distributed \emph{k}-d tree, then the complete set of squares is used to search the tree for intersecting rectangles. Because the squares themselves do not overlap, intersections between rectangles occur only within each square but do not extend to adjacent squares, so the expected number of intersecting rectangles is 9 times the number of squares.  This expected number is used to verify the correctness of the search result.

\begin{figure}[h]
\centering
\centerline{\includegraphics[width=3.5in]{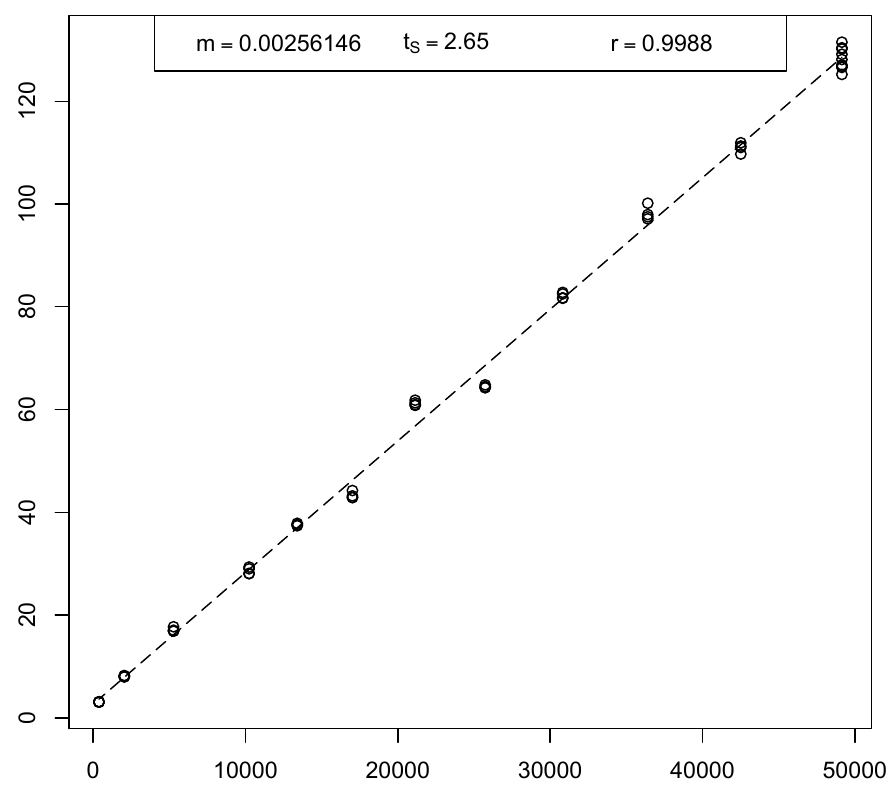}}
\caption{The \emph{k}-d tree building time (seconds) is plotted vs. $n \log_2 \left(n\right)$ for $2^{4} \le n \le 2^{12}$ rectangles.}
\label{fig:BuildTimeVsNlogN}
\end{figure}

Figure \ref{fig:BuildTimeVsNlogN} shows the execution time in seconds that is required by four workers to build a balanced \emph{k}-d tree, plotted versus $n \log_2 \left(n\right)$ for $2^{4} \le n \le 2^{12}$ rectangles. The dashed line of Figure \ref{fig:BuildTimeVsNlogN} shows the least-squares fit of the time $t$ to the function $t = mn \log_2 \left(n\right) + t_\mathrm{S}$ where $m$ is the slope of the line and the intercept $t_\mathrm{S}$ is a serial component of the execution time that is independent of $n$.  The correlation coefficient $r = 0.9988$ indicates an adequate least-squares fit; hence, the execution time is proportional to $n \log_2 \left(n\right)$. 

\begin{figure}[h]
\centering
\centerline{\includegraphics[width=3.5in]{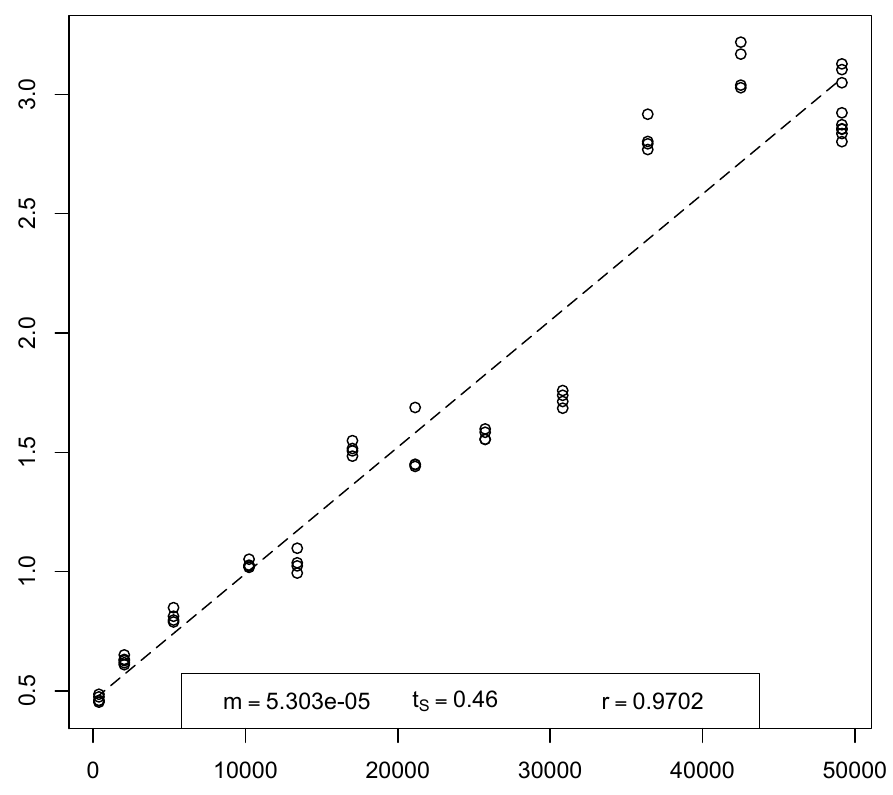}}
\caption{The \emph{k}-d tree search time (seconds) is plotted vs. $n \log_2 \left(n\right)$ for $2^{4} \le n \le 2^{12}$ rectangles.}
\label{fig:SearchTimeVsNlogN}
\end{figure}

Figure \ref{fig:SearchTimeVsNlogN} shows the execution time in seconds that is required by four workers to search the balanced \emph{k}-d tree for intersections between rectangles, plotted versus $n \log_2 \left(n\right)$ for $2^{4} \le n \le 2^{12}$ rectangles. The dashed line of Figure \ref{fig:SearchTimeVsNlogN} shows the least-squares fit of the time $t$ to the function $t = mn \log_2 \left(n\right) + t_\mathrm{S}$ where $m$ is the slope of the line and the intercept $t_\mathrm{S}$ is the serial component of the execution time.  The correlation coefficient $r = 0.9702$ indicates a less-than-adequate least-squares fit; hence, the execution time may not be strictly proportional to $n \log_2 \left(n\right)$.

In Figures \ref{fig:BuildTimeVsNlogN} and \ref{fig:SearchTimeVsNlogN}, the scales of the $y$-axes are 120 seconds and 3 seconds, respectively. Thus, the deviation from $O\left(n \log n\right)$ computational complexity that is apparent in Figure \ref{fig:SearchTimeVsNlogN} may be due to statistical variation that is present in both figures but more obvious in Figure \ref{fig:SearchTimeVsNlogN} than in Figure \ref{fig:BuildTimeVsNlogN}.

\begin{figure}[h]
\centering
\centerline{\includegraphics[width=3.5in]{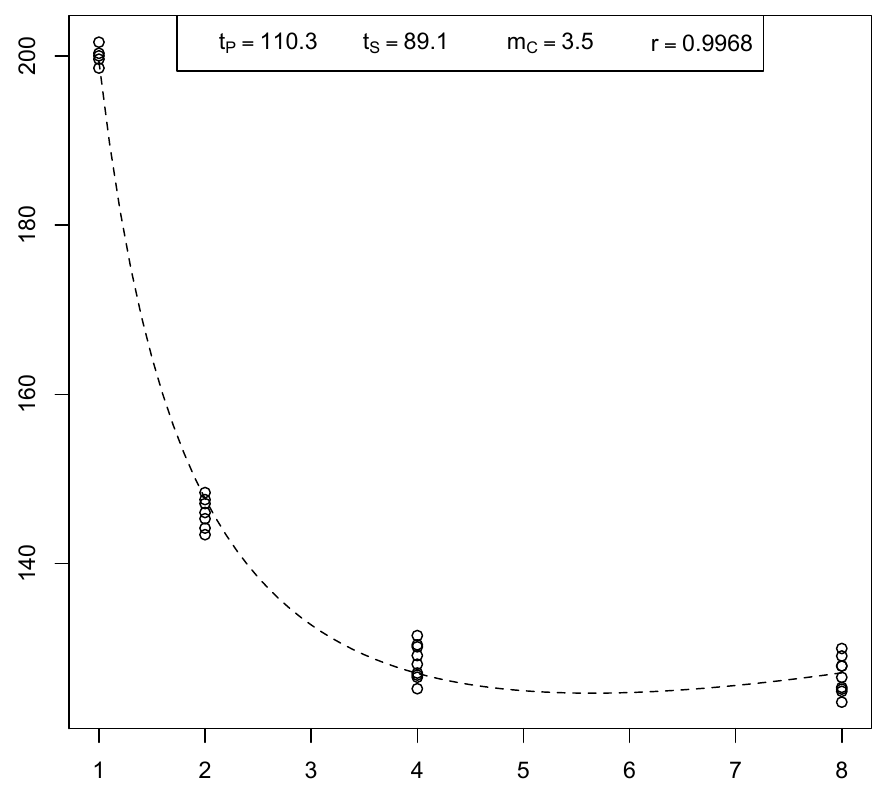}}
\caption{The \emph{k}-d tree building time (seconds) is plotted vs. the number of workers for $2^{12}$ rectangles.}
\label{fig:BuildParallel}
\end{figure}

\begin{figure}[h]
\centering
\centerline{\includegraphics[width=3.5in]{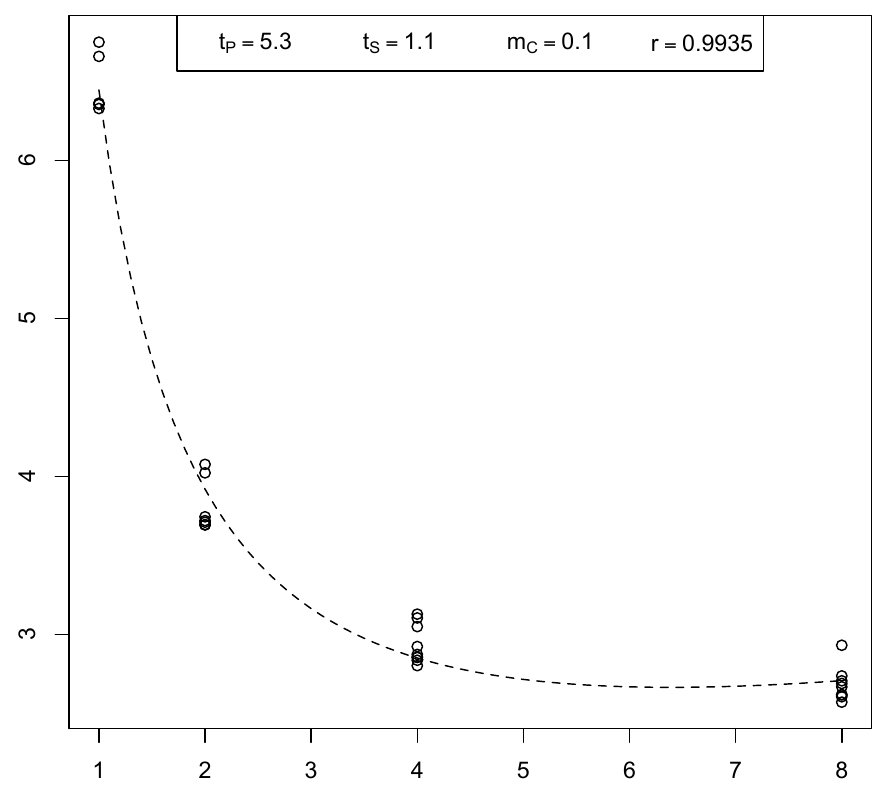}}
\caption{The \emph{k}-d tree search time (seconds) is plotted vs. the number of workers for $2^{12}$ rectangles.}
\label{fig:SearchParallel}
\end{figure}

The \emph{k}-d tree building and search algorithms were executed in parallel and their performance was measured for one to eight workers using a 2.3 GHz Intel quad-core i7 processor.  Figures \ref{fig:BuildParallel} and \ref{fig:SearchParallel} show, respectively, the execution time in seconds that is required to build and search the \emph{k}-d tree, plotted versus the number of workers $w$ for $n=2^{12}$ rectangles. The dashed curves in these figures show the least-squares fits of the execution time $t$ to the model
\begin{equation}
t =  t_\mathrm{s} + \frac{t_\mathrm{p}}{w} + m_\mathrm{c}\left(w - 1\right)
\label{eq:gunther}
\end{equation}
In this equation, $w$ is the number of workers, $t_\mathrm{s}$ represents the time required to execute the serial or non-parallelizable portion of the algorithm, $t_\mathrm{p}$ represents the time required to execute the parallelizable portion of the algorithm via one worker, and $m_\mathrm{c}\left(w - 1\right)$ models a limitation to parallel execution that the Amdahl model  \cite{Amdahl} fails to capture \cite{Gunther} and that likely arises due to cache misses \cite{Brown}. The correlation coefficients for the least-squares fits are 0.9968 and 0.9935 for Figures \ref{fig:BuildParallel} and \ref{fig:SearchParallel}, respectively, and indicate adequate least-squares fits.

\section{Discussion}
\label{sec:discussion}

Figure \ref{fig:BuildParallel} shows that one worker requires 200 seconds to build a distributed \emph{k}-d tree by subdividing sorted RDDs that contain $2^{12}$ rectangles.  In contrast, one worker requires only $122 \pm 3$ milliseconds to build a \emph{k}-d tree by subdividing sorted arrays that contain $2^{12}$ rectangles (based on 20 measurements of the execution time).

Given this disparity in performance, a reasonable strategy for building a distributed \emph{k}-d tree is to subdivide an RDD via workers but only to the point that each resulting subdivided RDD becomes small enough to fit into the memory of a compute node.  At that point, the \emph{k}-d tree building algorithm stops subdividing RDDs and begins subdividing arrays as follows.

A subdivided RDD contains all of the rectangles that are required for building a subtree, so the subdivided RDD is converted to an array that is returned to the master via the Spark collect method \cite{KarauD}.  Then the master constructs a subtree by subdividing the array.  Each node of the resulting subtree is then appended to the \textbf{tree RDD}.

Potentially, multiple subdivided RDDs would be produced, each of which would be converted to an array that is small enough to fit into the memory of a compute node.  Therefore, it is tempting to consider an execution model wherein the master would send an array representing the subdivided RDD to a worker via a remote procedure call (RPC).  The worker would build a subtree by subdividing the array then return the nodes of the subtree to the master that would append the nodes to the \textbf{tree RDD}.  Multiple workers would be available, so the master could send different arrays to different workers that could build different subtrees in parallel.

Because, for the performance measurements that are reported above, the master and workers comprise separate cores of an Intel quad-core i7 CPU, an effect similar to that of multiple RPCs may be achieved via execution of multiple threads that are provided by a Scala ExecutionContext.  This approach was implemented and shows that the execution time required to build a \emph{k}-d tree is independent of the number of threads dedicated to building subtrees by subdividing arrays.  This result is consistent with the fact that building a tree from $2^{12}$ rectangles by subdividing arrays occurs several thousand times more rapidly than building that tree by subdividing RDDs.  Hence, because an array is subdivided simultaneously to the subdivision of an RDD, subdivision of the array has completed long before subdivision of the RDD.  However, it is not always the case that building a subtree by subdividing arrays occurs more rapidly than subdividing RDDs because the time required for either depends on the depth in the tree.

At any particular level of the nascent \emph{k}-d tree, the time required to execute the Spark filter and splitAt methods in order to subdivide an RDD is directly proportional to the number of elements in the RDD, as evidenced by the fact that the time required to build the tree is proportional to $n \log_2 n$ (see Figure \ref{fig:BuildTimeVsNlogN}).  The time required to execute these methods is inversely proportional to the number of workers because these methods necessitate no communication between workers.  Hence, the time $t$ required to subdivide the $x_\mathrm{min}$:$n_\mathrm{i}$-sorted and $y_\mathrm{min}$:$n_\mathrm{i}$-sorted RDDs may be modeled as
\begin{equation}
t = \frac{c_\mathrm{r}n}{2^dw}
\label{eq:rdd}
\end{equation}
In this equation, $n$ is the number of rectangles, $d$ is the depth in the tree (where $d=0$ at the root), $w$ is the number of workers and $c_\mathrm{r}$ is a proportionality constant that is equal to the time required to build a \emph{k}-d tree from $n$ rectangles by subdividing RDDs, divided by $n \log_2 n$.

At any particular level of the nascent tree, the time required to build a subtree via subdivision of arrays may be modeled as
\begin{equation}
t = c_\mathrm{a} 2 \frac{n}{2^{d+1}} \log_2 \left( \frac{n}{2^{d+1}} \right)
\label{eq:mem}
\end{equation}
In this equation, $n$ is the number of rectangles, $d$ is the depth in the tree and $c_\mathrm{a}$ is a proportionality constant that is equal to the time required to build a \emph{k}-d tree from $n$ rectangles by subdividing arrays, divided by $n \log_2 n$.

The master can build a subtree via subdivision of arrays in less time than the time required for the workers to subdivide the $x_\mathrm{min}$:$n_\mathrm{i}$-sorted and $y_\mathrm{min}$:$n_\mathrm{i}$-sorted RDDs, so long as
\begin{equation}
c_\mathrm{a} 2 \frac{n}{2^{d+1}} \log_2 \left( \frac{n}{2^{d+1}} \right) < \frac{c_\mathrm{r}n}{2^dw}
\label{eq:compare}
\end{equation}
This equation can be simplified to obtain
\begin{equation}
d > \log_2 \left( n \right) - \frac{c_\mathrm{r}}{c_\mathrm{a}w} - 1
\label{eq:depth}
\end{equation}

The physical interpretation of Equation \ref{eq:depth} is that for a given number of rectangles $n$, increasing the ratio $c_\mathrm{r} / c_\mathrm{a} w$ means that for a smaller depth $d$ in the tree, the master can build a subtree via subdivision of arrays in less time than the time required for the workers to subdivide the $x_\mathrm{min}$:$n_\mathrm{i}$-sorted and $y_\mathrm{min}$:$n_\mathrm{i}$-sorted RDDs.

The \emph{k}-d tree building algorithm that was discussed in Sections \ref{sec:build_memory_kd_tree} and \ref{sec:build_distributed_kd_tree} of this article is not suitable for building a \emph{k}-d tree wherein RDDs are collected to arrays that are subsequently processed by asynchronous threads or RPCs.  That algorithm creates a bounding region as the recursion unwinds and returns the bounding region to the next higher level of the tree.  However, when the master launches an asynchronous thread to build a subtree, that thread cannot return the bounding region of the subtree to the master because the master does not wait for the thread to finish building the subtree.  Instead, the master subdivides the next pair of $x_\mathrm{min}$:$n_\mathrm{i}$-sorted and $y_\mathrm{min}$:$n_\mathrm{i}$-sorted RDDs while the thread builds the subtree.

For this reason, an alternate \emph{k}-d tree building algorithm is implemented for building the upper portion of the nascent tree via subdivision of RDDs.  This algorithm subdivides $x_\mathrm{min}$:$n_\mathrm{i}$-sorted, $y_\mathrm{min}$:$n_\mathrm{i}$-sorted, $x_\mathrm{max}$:$n_\mathrm{i}$-sorted and $y_\mathrm{max}$:$n_\mathrm{i}$-sorted RDDs.  It is possible to obtain the bounding region for a node in the \emph{k}-d tree from these four RDDs.  Specifically, for a given level of subdivision, the bounding region is constructed from the first element of the $x_\mathrm{min}$:$n_\mathrm{i}$-sorted RDD, the first element of the $y_\mathrm{min}$:$n_\mathrm{i}$-sorted RDD, the last element of the $x_\mathrm{max}$:$n_\mathrm{i}$-sorted RDD and the last element of the $y_\mathrm{max}$:$n_\mathrm{i}$-sorted RDD.

Because it is possible to construct the bounding region for a node from these four RDDs, there is no requirement to return the bounding region of a node as the recursion unwinds.  Hence, this algorithm for building a \emph{k}-d tree is well-suited to building the upper portion of the nascent tree.  For the lower portion of the nascent tree that is build via subdivision of arrays, the algorithm that was discussed in Sections \ref{sec:build_memory_kd_tree} and \ref{sec:build_distributed_kd_tree} of this article suffices.

 \vspace{0.25in}
 
\section{Conclusion}
\label{sec:conclusion}

This article presents a new \emph{k}-d tree building algorithm that builds a tree with MapReduce in time proportional to $n \log n$.  The tree is represented as a distributed graph that may be searched via a new \emph{k}-d tree search algorithm with MapReduce, possibly in time proportional to $n \log n$.

\subsection*{Source Code}

The \emph{k}-d tree building and search algorithms have been implemented in Scala and execute in parallel via Spark.  The source code for this implementation includes the BSD 3-Clause License and is available for download.

\subsection*{Acknowledgements}

The author thanks William Tyler and Joseph Wearing for helpful comments.
 
\bibliographystyle{abbrv}
\bibliography{building_distributed_kd_tree.bib}

\end{document}